\documentclass{Interspeech2024}

\interspeechcameraready
\usepackage{amsmath} 
\usepackage{algorithm}
\usepackage{algpseudocode}
\usepackage{xcolor}
\usepackage{multirow}
\algrenewcommand\algorithmicrequire{\textbf{Input:}}
\algrenewcommand\algorithmicensure{\textbf{Output:}}
\title{RawBMamba: End-to-End Bidirectional State Space Model for Audio Deepfake Detection}

\name[affiliation={1}]{Yujie}{Chen}
\name[affiliation={2, *}]{Jiangyan}{Yi} 
\name[affiliation={1}]{Jun}{Xue}
\name[affiliation={2}]{Chenglong}{Wang}
\name[affiliation={2}]{Xiaohui}{Zhang}
\name[affiliation={1}]{Shunbo}{Dong}
\name[affiliation={2}]{Siding}{Zeng}
\name[affiliation={3}]{Jianhua}{Tao}
\name[affiliation={1}]{Lv}{Zhao}
\name[affiliation={1}]{Cunhang}{Fan}

\address{
  $^1$ School of Computer Science and Technology, Anhui University, China 
  $^2$ Institute of Automation, Chinese Academy of Sciences, China
  $^3$ Department of Automation, Tsinghua University, China}

\email{e22201148@stu.ahu.edu.cn, jiangyan.yi@nlpr.ia.ac.cn}


\begin{document}

\maketitle
\renewcommand{\thefootnote}{\fnsymbol{footnote}}
\footnotetext{* Corresponding author.}
\setcounter{footnote}{0}
\renewcommand{\thefootnote}{\arabic{footnote}}
\begin{abstract}
Fake artefacts for discriminating between bonafide and fake audio can exist in both short-  and long-range segments. Therefore, combining local and global feature information can effectively discriminate between bonafide and fake audio. This paper proposes an end-to-end bidirectional state space model, named RawBMamba, to capture both short- and long-range discriminative information for audio deepfake detection. Specifically, we use sinc Layer and multiple convolutional layers to capture short-range features, and then design a bidirectional Mamba to address Mamba's unidirectional modelling problem and further capture long-range feature information. Moreover, we develop a bidirectional fusion module to integrate embeddings, enhancing audio context representation and combining short- and long-range information. The results show that our proposed RawBMamba achieves a 34.1\% improvement over Rawformer on ASVspoof2021 LA dataset, and demonstrates competitive performance on other datasets. Codes will be released on \url{https://github.com/cyjie429/RawBMamba.}
\end{abstract}

\begin{figure*}[t]
  \centering
  \includegraphics[width=\linewidth]{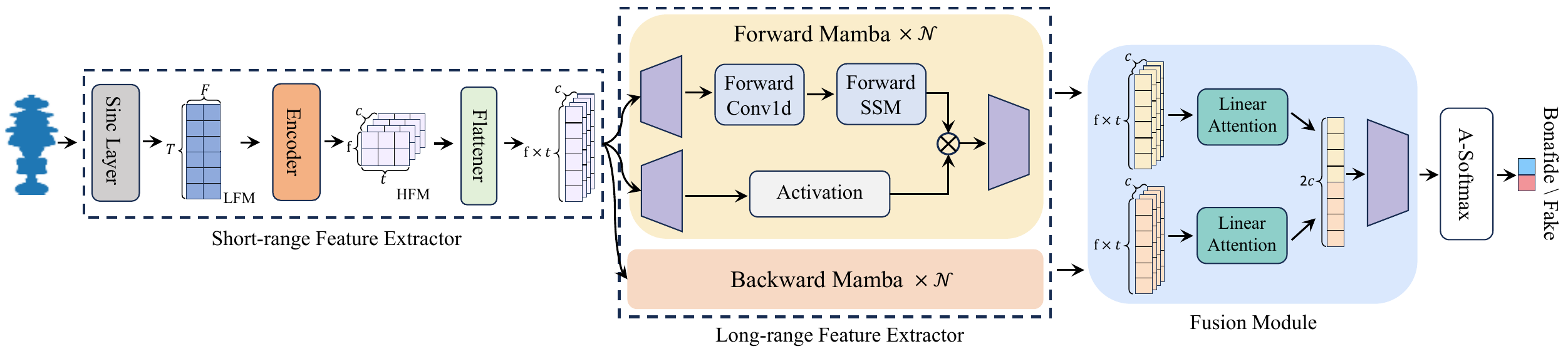}
  \caption{The overall structure diagram of our proposed RawBMamba.}
  \label{fig:RawBMamba}
\end{figure*}

\section{Introduction}

Recently, advances in deep learning have significantly accelerated the development of text-to-speech (TTS) and voice conversion (VC) technologies, making synthesised speech increasingly indistinguishable from real human voices. Nevertheless, the potential misuse of such technologies by malicious actors inevitably poses a significant threat to society. Hence, in recent years, a series of fake audio detection challenges \cite{nautsch2021asvspoof, yamagishi2021asvspoof, yi2022add, yi2023add} emerge with the aim of countering and bolstering the ability to detect fake audio attacks.

Extensive research demonstrates that discriminative information (i.e., fake artefacts) of synthesized and converted speech can be analyzed from both short-range and long-range segments of speech signals \cite{jung2022aasist, liu2023leveraging}. Abnormal intonation changes and unnatural stresses may exist in short-range segments, while unnatural rhythmic patterns and monotonous emotional expressions as deceptive cues may be present in long-range segments. Therefore, integrating both short-range and long-range information effectively enhances the system's detection capability. In recent years, mainstream audio deepfake detection approaches have been divided into pipeline \cite{xue2023learning, wang23x_interspeech, DBLP:conf/icml/ZhangYTWZ23, DBLP:conf/aaai/ZhangYWZZ024, fan2024dual} and end-to-end methods. The end-to-end model, by not relying on manually extracted features and directly optimizing on the raw audio waveform within a single model, can effectively enhance the model's generalizability \cite{Zhang_2024_CVPR}. Inspired by this, we believe that extracting both short- and long-range feature information from audio signals through an end-to-end approach can ensure the model's effectiveness and generalizability. Currently, extensive research focuses on end-to-end models\cite{zhang2024meraw}, RawNet2 \cite{tak2021end, hansen22_interspeech} by performing time-domain convolution operations directly on raw audio, possesses the potential to learn clues undetectable by knowledge-based methods. Additionally, researchers incorporate attention mechanisms into graph neural networks to capture key information across time and frequency \cite{DBLP:conf/interspeech/TakJ0TE21, DBLP:conf/iclr/VelickovicCCRLB18, tak2021end_gra}. AASIST \cite{jung2022aasist} captures complex short-range and long-range feature information in graph networks by modeling the non-Euclidean relationships of graph nodes. Rawformer \cite{liu2023leveraging} integrates convolutional layers with Transformer \cite{vaswani2017attention} structures, utilizing self-attention to capture short-range and long-range feature information in speech signals. Although the aforementioned end-to-end models are already capable of handling audio deepfake detection, there still exists room for improvement \cite{DBLP:conf/dada/ZhangYTWXF23}.

Recently, inspired by the success of the state space model with efficient hardware-aware designs (i.e., Mamba \cite{gu2023mamba}) in the field of sequence modeling, which selects relevant information to model long-range relationships, empirical evidence has demonstrated its enhanced capability in capturing long-range relations more effectively. Therefore, we believe it can also more effectively capture long-range feature information in speech signals. Based on this, we aim to design an efficient end-to-end model that integrates both short-range and long-range feature information. However, Mamba, due to its Markovian nature, suffers from the drawback of unidirectional modeling, which prevents it from fully capturing contextual information, thereby affecting the acquisition of long-range information.

To address these limitations, we propose a End-to-End Bidirectional Mamba model, named RawBMamba, to more effectively capture short- to long-range feature information. Specifically, we utilize a series of parametrizable sinc functions to obtain low-level feature mappings, followed by acquiring high-level feature mappings through multiple convolutional blocks, followed by mapping reconstruction to generate a two-dimensional sequence suitable for bidirectional Mamba. The Mamba models are then used to capture forward and backward long-range features respectively. Finally, a bidirectional feature fusion module merges the two sets of embeddings to obtain the comprehensive short-range to long-range feature representation. In addition, we perform a preliminary theoretical analysis of the effectiveness of Mamba and Transformers in the domain of audio deepfake detection to demonstrate the viability of Mamba. We achieve competitive experimental results on multiple datasets, proving the effectiveness and generalization ability of RawBMamba. 

\section{Preliminaries}
Mamba is inspired by continuous state space models (SSM) \cite{gu2021combining, gu2022parameterization, gupta2022diagonal, DBLP:conf/iclr/GuGR22} in control systems. It maps a 1-D function or sequence $ x(t) \in \mathbb{R} \rightarrow y(t) \in \mathbb{R} $ through a hidden state $ h(t) \in \mathbb{R}^N $. This model uses $ \mathbf{A} \in \mathbb{R} ^ {N \times N} $ as the State Transition parameters, and $ \mathbf{B} \in \mathbb{R} ^ {N \times 1} $, $ \mathbf{C} \in \mathbb{R} ^ {1 \times N} $ as the projection parameters.
\begin{equation}
\begin{split}
h'(t) &= \mathbf{A}h(t) + \mathbf{B}x(t), \\
y(t) &= \mathbf{C}h(t).
\end{split}
\end{equation}
Mamba is the classic discrete versions of the continuous system. it include a timescale parameter $\mathbf{\Delta}$, and through discretization rules, transform the continuous parameters $\mathbf{A}$, $\mathbf{B}$ to discrete parameters $\overline{\mathbf{A}}$, $\overline{\mathbf{B}}$.A commonly used discretization rule is the zero-order hold (ZOH), which is defined as follows:
\begin{equation}
\begin{split}
\overline{\mathbf{A}} &= \exp(\mathbf{\Delta} \mathbf{A}), \\
\overline{\mathbf{B}} &= (\mathbf{\Delta} \mathbf{A})^{-1}(\exp(\mathbf{\Delta} \mathbf{A}) - \mathbf{I}) \cdot \Delta \mathbf{B}.
\end{split}
\end{equation}
After the discretization of $\overline{\mathbf{A}}$, $\overline{\mathbf{B}}$, the discretized version of Eq. (1) using a step size $\mathbf{\Delta}$ can be rewritten as:
\begin{equation}
\begin{split}
h_t &= \overline{\mathbf{A}}h_{t-1} + \overline{\mathbf{B}}x_t, \\
y_t &= \mathbf{C}h_t.
\end{split}
\end{equation}
At last, the models compute output through a global convolution.
\begin{equation}
\begin{split}
\overline{\mathbf{K}} &= (\mathbf{C}\overline{\mathbf{B}}, \mathbf{C}\overline{\mathbf{A}}\overline{\mathbf{B}}, \ldots, \mathbf{C}\overline{\mathbf{A}}^{M-1}\overline{\mathbf{B}}), \\
y &= x * \overline{\mathbf{K}}.
\end{split}
\end{equation}
where M is the length of the input sequence $\mathbf{x}$, and $\overline{\mathbf{K}} \in \mathbb{R}^M$ is a structured convolutional kernel.

\section{Proposed Method}
In this section, we provide a detailed introduction to RawBMamba, as shown in Figure \ref{fig:RawBMamba} . First,  we describe how RawBMamba captures short-range feature representation. Then, we present the implementation details of the Bidirectional Mamba and the Bidirectional Feature Fusion Module.

\subsection{Short-range feature representation}

An increasing number of researchers are using trainable neural layers to learn approximate standard filter processes from raw waveforms. RawBMamba employs a variant of RawNet2's frontend to capture high-level short-range feature mappings (HFM). Specifically, first, a series of parametric sinc functions are utilized to implement band-pass filters \cite{ravanelli2018speaker}, extracting spectral-temporal features from the raw waveform, forming Low-level Feature Maps (LFM) $F_{LFM} \in \mathbb{R}^{F \times T}$, where where $F$ and $T$ are the numbers of frequency and the temporal bins, respectively. Then, High-level Feature Maps (HFM) $F_{HFM} \in \mathbb{R}^{C \times f \times t}$ containing short-range feature information are extracted through a series of three ResNet blocks with squeeze-and-excitation operations, here, $C$, $f$ and $t$ denote the number of channels, frequency bins, and temporal locations after dimensionality reduction. Finally, the high-level feature maps are flattened along the time and frequency axes to obtain a two-dimensional short-range feature sequence $F_{s} \in \mathbb{R}^{C \times ft}$, suitable as input vectors for the bidirectional Mamba.

\begin{algorithm}[t]
\caption{Bidirectional Mamba}
\label{alg:algorithm1}
\begin{algorithmic} %
\Require token sequence $F_{S}: (B, L, C)$
\Ensure token sequence $F_{forward}, F_{backward}: (B, L, C)$
\State \textcolor{gray}{/* sequence reverse operation  */}
\State $F_{forward}: (B, L, C) \gets F_{short}$ 
\State $F_{backward}: (B, L, C) \gets \text{Reverse}(F_{short})$ 
\State \textcolor{gray}{/* process with different direction */}
\For{d in \{forward, backward\}}
    \State $x: (B, L, C) \gets F_d$
    \For{i in $Mamba Layer_d$}
        \State $x_i, z: (B, L, E) \gets \text{Linear}^{xz}(x))$
        \State $x'_i: (B, L, E) \gets \text{SiLU}(\text{Conv1d}(x_i))$
        \State \textcolor{gray}{/* Selection Mechanism */}
        \State $\mathbf{B}_i: (B, L, N) \gets \text{Linear}^B_i(x'_i)$
        \State $\mathbf{C}_i: (B, L, N) \gets \text{Linear}^C_i(x'_i)$
        \State \textcolor{gray}{/* softplus ensures positive $\mathbf{\Delta}_i$ */ }
        \State $\mathbf{\Delta}_i: (B, L, E) \gets \log(1 + \exp(\text{Linear}^{\Delta}_i(x'_{i}) + \text{Parameter}^{\Delta}_{i}))$
        \State \textcolor{gray}{/* shape of $\text{Parameter}^{\mathbf{A}}_{i}$ is $(E, N)$ */}
        \State $\overline{\mathbf{A}_i}: (B, L, E, N) \gets \mathbf{\Delta}_i \otimes \text{Parameter}^{\mathbf{A}}_{i}$
        \State $\overline{\mathbf{B}_i}: (B, L, E, N) \gets \mathbf{\Delta}_i \otimes \mathbf{B}_i$
        \State $y_i: (B, L, C) \gets \text{Linear}(\text{SSM}(\overline{\mathbf{A_i}}) \otimes \text{SiLU}(z))$
        \State $x: (B, L, C) \gets y_i$
    \EndFor
    \State $F_{d}: (B, L, C) \gets y_i$
\EndFor
\State return $F_{forward}, F_{backward}$
\end{algorithmic}
\end{algorithm}

\subsection{Bidirectional state space model}
Mamba designs a simple selection mechanism that, by parameterizing the sequence dimension, enables the model to effectively select the most relevant information and filter out irrelevant information, thereby enhancing the model's ability to capture long-range features. However, due to its Markovian nature, it can only model long-range features in a unidirectional manner, which leads to ineffective capture of contextual information. Therefore, we design the bidirectional Mamba, as illustrated in Figure \ref{fig:RawBMamba}, which specifically consists of the following two parts:

\textbf{Bidirectional Mamba.} We present Algorithm \ref{alg:algorithm1} for the bidirectional Mamba. we first take the short-range features $F_{s}$ obtained from the frontend and generate a backward short-range feature through a reverse sequence operation. Then, we deploy two structurally identical Mamba networks to independently capture the long-range feature information of the forward and backward short-range feature. Specifically, we project the forward and backward features onto $x_i$ and $z$ with dimensions of $E$, respectively. Then a 1-D convolution is applied to $x$ to obtain $x'_i$. Following that, $x'_i$ is projected to obtain $\mathbf{B}_i$, $\mathbf{C}_i$ and $\mathbf{\Delta}_i$. Discretisation rules are used to generate $\overline{\mathbf{A}_i}$, $\overline{\mathbf{B}_i}$. Finally, the forward long-range features $F_{forward}$ and the backward long-range features $F_{backward}$ are obtained.

\textbf{Bidirectional feature fusion block.} After acquiring the forward and backward long-range features, we initially apply a linear self-attention operation on each of the unidirectional long-range features to extract key information. Subsequently, we employ concatenation for bidirectional feature fusion. Finally, we obtain short- and long-range speech features $F_l$ enriched with contextual information, which are used for authenticity discrimination. The specific formula is illustrated as follows:
\begin{equation}
F_1, F_2 = Attention(F_{forward}), Attention(F_{backward})
\end{equation}
\begin{equation}
F_l = MLP(Concat(F_1, F_2))
\end{equation}

\section{Experiments and Analysis}

\subsection{Datasets and metrics}
We conduct extensive tests on the ASVspoof2019 LA (19LA), ASVspoof2021 LA (21LA) and ASVspoof2021 DF (21DF) to evaluate our model's effectiveness and generalizability. 19LA features three types of fake attacks (TTS, VC) across 19 algorithms (A01-A19). 21LA contains real and fake speeches via telephony systems like Voice over IP and Public Switched Telephone Network. 21DF comprises authentic and spoofed voices altered by various media codecs, which introduces distortion during the encoding, compressing, and subsequent decoding of audio data. The performance assessment uses the equal error rate (EER) and the minimum tandem detection cost function (min t-DCF). 

\subsection{Implementation details}
In the training phase, we use input data with 64,000 sampling points ($ \approx 4 $ seconds) for RawBMamba. The SincNet layer is configured with 70 filters, and the variant based on RawNet2 consists of four sub-blocks, with the first two having 32 filters and the latter two having 64 filters each. We utilize an Adam optimizer with a learning rate of $1 \times 10^{-5}$, and training batch size is set to 32. The model is trained on the ASVspoof 2019 LA training and development sets for 32 epochs using the A-Softmax\cite{liu2017sphereface} loss function on a single RTX 3090 GPU. 

\begin{table}[]
\centering
\caption{The experimental results of RawBMamba with different configurations on the 19LA, 21LA, and 21DF datasets. Here, "uni" is the unidirectional Mamba and "bi" is the bidirectional Mamba. "L" is model's total layers.}
\begin{tabular}{@{}ccc@{\hspace{0.3em}}cc@{\hspace{0.3em}}cc@{}}
\toprule
\multirow{2}{*}{\textbf{Dir.}} & \multirow{2}{*}{\textbf{L}} & \multicolumn{2}{c}{\textbf{19LA}} & \multicolumn{2}{c}{\textbf{21LA}} & \textbf{21DF} \\ 
\cmidrule(r){3-7}
                                    &                                 & \textbf{EER(\%)}       & \textbf{t-DCF}       & \textbf{EER(\%)}       & \textbf{t-DCF}       & \textbf{EER(\%)}          \\ \midrule
\multirow{3}{*}{uni}    & 4   & 2.51   & 0.0707   & 3.74   & 0.2707   & \textbf{20.94}                     \\ 
                        & 8   & 2.01   & 0.0623   & 3.44   & 0.2652   & 21.22                     \\ 
                        & 12  & \textbf{1.47}   & \textbf{0.0467}   & \textbf{2.84}   & \textbf{0.2517}   & 22.48                     \\ \midrule
\multirow{3}{*}{bi}     & 4   & \textbf{1.07}   & \textbf{0.0315}   & 3.38   & 0.2687   & 20.83                          \\  
                        & 8   & 1.67   & 0.0484   & 3.45   & \textbf{0.2631}   & 18.80                      \\ 
                        & 12   & 1.19   & 0.0360   & \textbf{3.28}   & 0.2709   & \textbf{15.85}                     \\ 
\bottomrule
\end{tabular}
\label{tab:Table 1}
\end{table}

\begin{table}[]
\centering
\caption{The experimental results of RawBMamba with different fusion methods on the 19LA, 21LA, and 21DF datasets.}
\begin{tabular}{@{}cc@{\hspace{0.5em}}cc@{\hspace{0.5em}}cc@{}}
\toprule
\multirow{2}{*}{\textbf{Methods}} & \multicolumn{2}{c}{\textbf{19LA}} & \multicolumn{2}{c}{\textbf{21LA}} & \textbf{21DF} \\
\cmidrule(r){2-6}
& \textbf{EER(\%)}       & \textbf{t-DCF}       & \textbf{EER(\%)}       & \textbf{t-DCF}       & \textbf{EER(\%)}          \\ 
\midrule
Sum  & 1.27 & 0.0400 & 4.13 & 0.2924 & 16.58 \\ 
Concat  & \textbf{1.19} & \textbf{0.0360} & 3.28 & 0.2709 & \textbf{15.85} \\
Attention  & 1.19 & 0.0369 & \textbf{3.19} & \textbf{0.2620} & 18.42 \\ 
\bottomrule
\end{tabular}
\label{tab:Table 2}
\end{table}

\subsection{Results of RawBMamba with different configurations}
This section discusses the unidirectional and bidirectional Mamba models' capability to capture short- and long-range features with \(N\) layers, as detailed in Table \ref{tab:Table 1}. We observe that the unidirectional Mamba model struggles with the 21DF dataset and only shows effectiveness on 21LA, indicating limitations in capturing generalized long-range features. However, when analysing comprehensive results from all datasets, the bidirectional Mamba model consistently outperforms the unidirectional Mamba model. This shows that bidirectional Mamba can overcome the problem of insufficient contextual information capture in unidirectional Mamba, thereby improving the generalizability of long-range features. In addition, we observe that the bidirectional Mamba model maintains good performance even with a lower number of layers (i.e., \(N=4\)) compared to unidirectional Mamba models with fewer layers, which we attribute to the effectiveness of our bidirectional feature fusion module in fully exploiting information from both forward and backward long-range features. However, the reduction in the number of layers inevitably leads to a reduction in the ability to handle difficult datasets such as 21DF.

\subsection{Results of RawBMamba with different fusion methods}
In this section, we will discuss in detail the impact of different fusion methods in the bidirectional feature fusion module of RawBMamba, as shown in Table \ref{tab:Table 2}. Specifically, we  design three common fusion methods: summation, concatenation, and the cross-attention mechanism. Both summation and cross-attention mechanism act directly on the two unidirectional long-range features. The results show that concatenation achieves the best overall effect compared to the other two fusion methods, maintaining good performance across all datasets. This suggests that while preserving the integrity of the time-frequency information as much as possible, it is able to capture the key information of the two unidirectional long-range features in the time-frequency domain. The cross-attention mechanism performs poorly on the 21DF dataset, suggesting that it may overfocus on key information and lose time-frequency information, reducing generalisation to out-of-domain data. In contrast, the summation fusion method shows promising results on the 21DF dataset, as it involves a simple summation operation on all time-frequency domain information of the two unidirectional long-range features, acting as a potential regularization effect that balances the information across all time-frequency domains.

\begin{table}[]
\centering
\caption{The comparison results of RawBMamba, Rawformer, and AASIST on the 19LA, 21LA, and 21DF datasets. Here, "$\dagger$" is reproduced results.}
\begin{tabular}{@{}c@{\hspace{0.3em}}c@{\hspace{0.3em}}c@{\hspace{0.5em}}c@{\hspace{0.3em}}c@{\hspace{0.5em}}c@{}}
\toprule
\multirow{2}{*}{\textbf{Models}} & \multicolumn{2}{c}{\textbf{19LA}} & \multicolumn{2}{c}{\textbf{21LA}} & \textbf{21DF} \\ \cmidrule(r){2-6}
                                 & \textbf{EER(\%)}       & \textbf{t-DCF}       & \textbf{EER(\%)}       & \textbf{t-DCF}       & \textbf{EER(\%)}          \\ 
\midrule
T23 \cite{yamagishi2021asvspoof}  & -                   & -               & 1.32                  & 0.2177               & 15.64        \\
RawNet2 \cite{tak2021end}  & -                   & -               & 9.50                  & -              & 22.38                         \\
TO-RawBet \cite{wang23w_interspeech}  & 1.58                   & -               & 3.70                 & -              & -                         \\
AASIST \cite{jung2022aasist}                           & \textbf{0.93}                   & \textbf{0.0285}               & 10.51                  & 0.4884               & -                         \\ 
AASIST-4Block \cite{jung2022aasist}                    & 1.20                    & 0.0341               & 9.15                   & 0.4370                & -                          \\
ARawNet2 \cite{li23h_interspeech}  & 4.61                    & -               & 8.36                   & -                & 19.03                          \\
SE-Rawformer \cite{liu2023leveraging}                     & 1.05                   & 0.0344               & 4.98                   & 0.3186               & -                         \\ 
SE-Rawformer$\dagger$                    & 1.15                   & 0.0314               & 4.31                   & 0.2851               & 20.26                     \\ 
\midrule
RawMamba(ours)                         & 1.47                   & 0.0467               & \textbf{2.84}                   & \textbf{0.2517}               & 22.48                     \\ 
\textbf{RawBMamba(ours)}                        & 1.19                   & 0.0360               & 3.28                    & 0.2709               & \textbf{15.85}                    \\ 
\bottomrule
\end{tabular}
\label{tab:Table 3}
\end{table}

\subsection{Comparison with the other end-to-end models}
In Table \ref{tab:Table 3}, we compare the performance of RawBMamba on 19LA, 21LA, and 21DF with the other end-to-end models. RawMamba refers to a 12-layer unidirectional Mamba model, whereas RawBMamba is a 12-layer bidirectional Mamba model, with 6 layers per direction.

To verify RawBMamba's generalization capability, We conduct extensive experiments on three datasets, with RawBMamba consistently performing well across all. Specifically, on the 21LA dataset, our proposed RawBMamba significantly outperforms the other two baselines. RawBMamba shows a 34.1\% performance improvement over SE-Rawformer, and RawMamba shows a 43.0\% improvement over SE-Rawformer, with comparable results on the 19LA dataset. This indicates that Mamba effectively captures long-range feature information more efficiently than Transformer. Furthermore, experimental results on the 21DF dataset show that RawBMamba, as a single system, achieves performance extremely close to that of the multi-system score fusion T23. This indicates that RawBMamba maintains commendable performance even when faced with complex, diverse, and unknown speech forgery methods. It demonstrates RawBMamba's robustness to out-of-domain data, further proving its generalization capability.

\subsection{Effectiveness analysis of RawBMamba and Rawformer}
\begin{figure}[t]
  \centering
  \includegraphics[width=\linewidth]{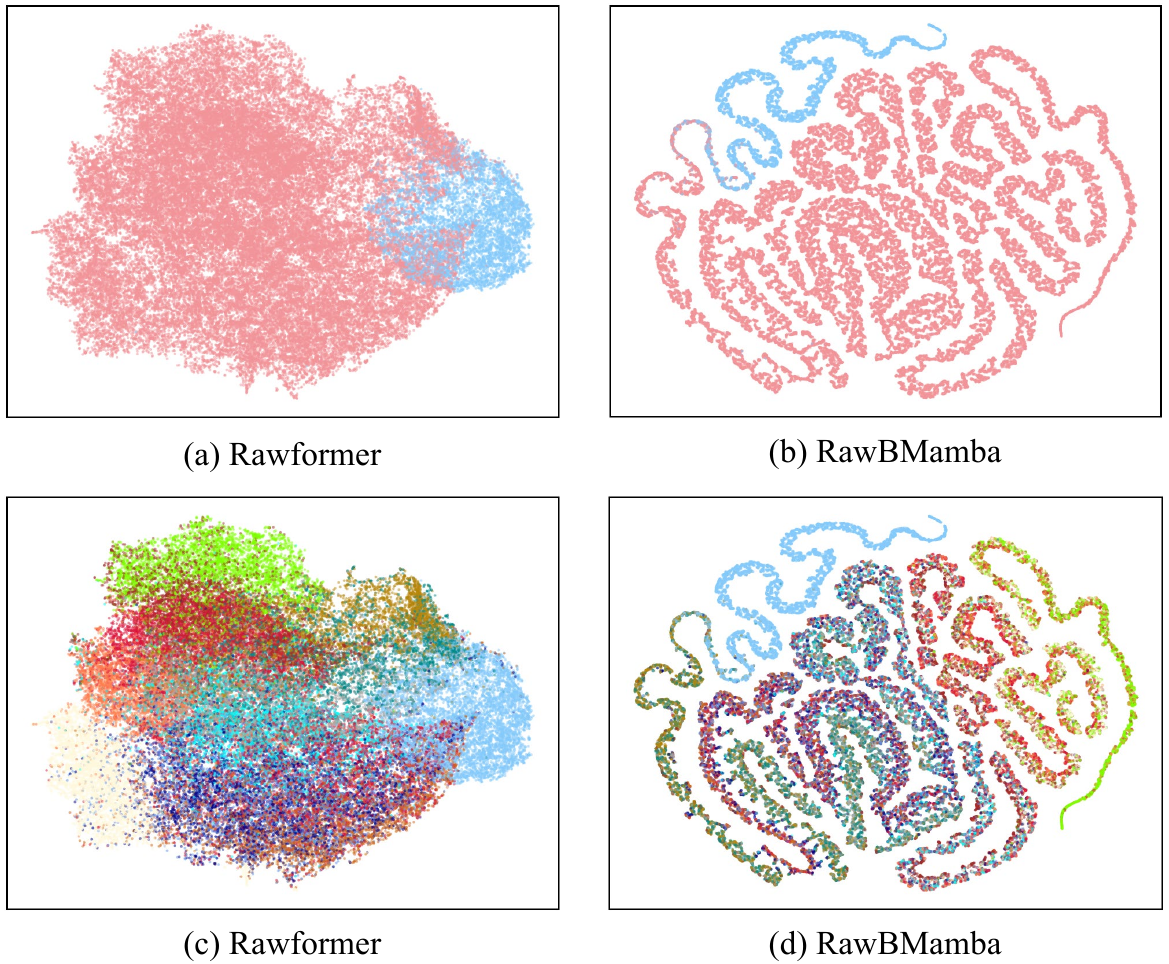}
  \caption{The 19LA test set samples' clustering is shown in 2D t-SNE plots from the model's higher layers. The visualization displays the clustering of bonafide vs. fake audio in (a) and (b) (blue for bonafide, red for fake) and different attack types in (c) and (d), with each color indicating a specific attack.}
  \label{fig:Transformer_Mamba}
\end{figure}
We discuss whether the Mamba architecture has more potential than the Transformer architecture for handling the audio fake detection task. Specifically, we directly extract the final features on the 19LA test dataset using the Rawformer model and the RawBMamba model. Then, we use t-SNE \cite{van2008visualizing}, a nonlinear dimensionality reduction algorithm for visualizing high-dimensional data, to visualize the features. The results are shown in Figure \ref{fig:Transformer_Mamba}. From (a) and (b), it is not difficult to see that when dealing with the binary task of bonafide and fake audio detection, there is partial block overlap between bonafide and fake audio for Transformer, while Mamba only has thread-like overlap at the end of the curve-shaped figures, indicating that the features obtained by Mamba are more discriminable than those extracted by Transformer. Furthermore, we also discover that Mamba's feature visualization presents multiple non-overlapping curvilinear patterns, embodying richer feature information capable of distinguishing different attack methods. This indicates Mamba architecture's potential in handling fine-grained classification tasks, as shown in (c) and (d). When mapping the sample points of different attack methods, the distribution of Mamba is clearer, which We believe is due to the key role of Mamba's sequence selection mechanism in obtaining discriminative information between different attack methods.

Therefore, we believe the Mamba architecture is capable of handling audio deepfake detection and has significant potential to become a backbone in audio deepfake detection models.

\section{Conclusion}
In this paper, we propose an end-to-end state space model, named RawBMamba, to capture both short- and long-range discriminative information in audio signals. Specifically, We use parametrizable sincLayers and multiple convolutional layers to capture short-range features, then design a bidirectional Mamba to address the limitations of Mamba's unidirectional modeling. We use a bidirectional feature fusion module to merge forward and backward long-range features, enhancing audio context representation. Experiments indicate that RawBMamba achieves a 34.1\% improvement over the Rawformer on ASVspoof 2021 LA and outperforms the state-of-the-art end-to-end models on ASVspoof 2021 LA, validating RawBMamba's generalizability and capability to handle out-of-domain data. Visualization analysis on t-SNE plots further validates the effectiveness of RawBMamba and its potential in fine-grained classification tasks. In future work, we plan to explore the application of RawBMamba to provenance tasks.

\section{Acknowledgements}
This work is supported by the Scientific and Technological Innovation Important Plan of China (No. 2021ZD0201502), the National Natural Science Foundation of China (NSFC) (No.62322120, No.U21B2010, No.62306316, No.62206278, No.62201002).

\bibliographystyle{IEEEtran}
\bibliography{mybib}

\begin{thebibliography}{10}
\providecommand{\url}[1]{#1}
\csname url@samestyle\endcsname
\providecommand{\newblock}{\relax}
\providecommand{\bibinfo}[2]{#2}
\providecommand{\BIBentrySTDinterwordspacing}{\spaceskip=0pt\relax}
\providecommand{\BIBentryALTinterwordstretchfactor}{4}
\providecommand{\BIBentryALTinterwordspacing}{\spaceskip=\fontdimen2\font plus
\BIBentryALTinterwordstretchfactor\fontdimen3\font minus \fontdimen4\font\relax}
\providecommand{\BIBforeignlanguage}[2]{{%
\expandafter\ifx\csname l@#1\endcsname\relax
\typeout{** WARNING: IEEEtran.bst: No hyphenation pattern has been}%
\typeout{** loaded for the language `#1'. Using the pattern for}%
\typeout{** the default language instead.}%
\else
\language=\csname l@#1\endcsname
\fi
#2}}
\providecommand{\BIBdecl}{\relax}
\BIBdecl

\bibitem{nautsch2021asvspoof}
A.~Nautsch, X.~Wang, N.~Evans, T.~H. Kinnunen, V.~Vestman, M.~Todisco, H.~Delgado, M.~Sahidullah, J.~Yamagishi, and K.~A. Lee, ``Asvspoof 2019: spoofing countermeasures for the detection of synthesized, converted and replayed speech,'' \emph{IEEE Transactions on Biometrics, Behavior, and Identity Science}, vol.~3, no.~2, pp. 252--265, 2021.

\bibitem{yamagishi2021asvspoof}
J.~Yamagishi, X.~Wang, M.~Todisco, M.~Sahidullah, J.~Patino, A.~Nautsch, X.~Liu, K.~A. Lee, T.~Kinnunen, N.~Evans \emph{et~al.}, ``Asvspoof 2021: accelerating progress in spoofed and deepfake speech detection,'' in \emph{ASVspoof 2021 Workshop-Automatic Speaker Verification and Spoofing Coutermeasures Challenge}, 2021.

\bibitem{yi2022add}
J.~Yi, R.~Fu, J.~Tao, S.~Nie, H.~Ma, C.~Wang, T.~Wang, Z.~Tian, Y.~Bai, C.~Fan \emph{et~al.}, ``Add 2022: the first audio deep synthesis detection challenge,'' in \emph{ICASSP 2022-2022 IEEE International Conference on Acoustics, Speech and Signal Processing (ICASSP)}.\hskip 1em plus 0.5em minus 0.4em\relax IEEE, 2022, pp. 9216--9220.

\bibitem{yi2023add}
J.~Yi, J.~Tao, R.~Fu, X.~Yan, C.~Wang, T.~Wang, C.~Y. Zhang, X.~Zhang, Y.~Zhao, Y.~Ren \emph{et~al.}, ``Add 2023: the second audio deepfake detection challenge,'' \emph{arXiv preprint arXiv:2305.13774}, 2023.

\bibitem{jung2022aasist}
J.-w. Jung, H.-S. Heo, H.~Tak, H.-j. Shim, J.~S. Chung, B.-J. Lee, H.-J. Yu, and N.~Evans, ``Aasist: Audio anti-spoofing using integrated spectro-temporal graph attention networks,'' in \emph{ICASSP 2022-2022 IEEE International Conference on Acoustics, Speech and Signal Processing (ICASSP)}.\hskip 1em plus 0.5em minus 0.4em\relax IEEE, 2022, pp. 6367--6371.

\bibitem{liu2023leveraging}
X.~Liu, M.~Liu, L.~Wang, K.~A. Lee, H.~Zhang, and J.~Dang, ``Leveraging positional-related local-global dependency for synthetic speech detection,'' in \emph{ICASSP 2023-2023 IEEE International Conference on Acoustics, Speech and Signal Processing (ICASSP)}.\hskip 1em plus 0.5em minus 0.4em\relax IEEE, 2023, pp. 1--5.

\bibitem{xue2023learning}
J.~Xue, C.~Fan, J.~Yi, C.~Wang, Z.~Wen, D.~Zhang, and Z.~Lv, ``Learning from yourself: A self-distillation method for fake speech detection,'' in \emph{ICASSP 2023-2023 IEEE International Conference on Acoustics, Speech and Signal Processing (ICASSP)}.\hskip 1em plus 0.5em minus 0.4em\relax IEEE, 2023, pp. 1--5.

\bibitem{wang23x_interspeech}
C.~Wang, J.~Yi, J.~Tao, C.~Y. Zhang, S.~Zhang, and X.~Chen, ``{Detection of Cross-Dataset Fake Audio Based on Prosodic and Pronunciation Features},'' in \emph{Proc. INTERSPEECH 2023}, 2023, pp. 3844--3848.

\bibitem{DBLP:conf/icml/ZhangYTWZ23}
X.~Zhang, J.~Yi, J.~Tao, C.~Wang, and C.~Y. Zhang, ``Do you remember? overcoming catastrophic forgetting for fake audio detection,'' in \emph{International Conference on Machine Learning, {ICML} 2023}, ser. Proceedings of Machine Learning Research, vol. 202.\hskip 1em plus 0.5em minus 0.4em\relax {PMLR}, 2023, pp. 41\,819--41\,831.

\bibitem{DBLP:conf/aaai/ZhangYWZZ024}
X.~Zhang, J.~Yi, C.~Wang, C.~Y. Zhang, S.~Zeng, and J.~Tao, ``What to remember: Self-adaptive continual learning for audio deepfake detection,'' in \emph{Thirty-Eighth {AAAI} Conference on Artificial Intelligence, {AAAI} 2024}, M.~J. Wooldridge, J.~G. Dy, and S.~Natarajan, Eds.\hskip 1em plus 0.5em minus 0.4em\relax {AAAI} Press, 2024, pp. 19\,569--19\,577.

\bibitem{fan2024dual}
C.~Fan, M.~Ding, J.~Tao, R.~Fu, J.~Yi, Z.~Wen, and Z.~Lv, ``Dual-branch knowledge distillation for noise-robust synthetic speech detection,'' \emph{IEEE/ACM Transactions on Audio, Speech, and Language Processing}, 2024.

\bibitem{Zhang_2024_CVPR}
X.~Zhang, J.~Yoon, M.~Bansal, and H.~Yao, ``Multimodal representation learning by alternating unimodal adaptation,'' in \emph{Proceedings of the IEEE/CVF Conference on Computer Vision and Pattern Recognition (CVPR)}, June 2024, pp. 27\,456--27\,466.

\bibitem{zhang2024meraw}
X.~Zhang, W.~Fu, and M.~Liang, ``Multimodal emotion recognition from raw audio with sinc-convolution,'' \emph{arXiv preprint arXiv:2402.11954}, 2024.

\bibitem{tak2021end}
H.~Tak, J.~Patino, M.~Todisco, A.~Nautsch, N.~Evans, and A.~Larcher, ``End-to-end anti-spoofing with rawnet2,'' in \emph{ICASSP 2021-2021 IEEE International Conference on Acoustics, Speech and Signal Processing (ICASSP)}.\hskip 1em plus 0.5em minus 0.4em\relax IEEE, 2021, pp. 6369--6373.

\bibitem{hansen22_interspeech}
J.~H. Hansen and Z.~WANG, ``Audio anti-spoofing using simple attention module and joint optimization based on additive angular margin loss and meta-learning,'' in \emph{Proc. Interspeech 2022}, 2022, pp. 376--380.

\bibitem{DBLP:conf/interspeech/TakJ0TE21}
H.~Tak, J.~Jung, J.~Patino, M.~Todisco, and N.~W.~D. Evans, ``Graph attention networks for anti-spoofing,'' in \emph{Interspeech 2021, 22nd Annual Conference of the International Speech Communication Association, Brno, Czechia, 30 August - 3 September 2021}.\hskip 1em plus 0.5em minus 0.4em\relax {ISCA}, 2021, pp. 2356--2360.

\bibitem{DBLP:conf/iclr/VelickovicCCRLB18}
P.~Velickovic, G.~Cucurull, A.~Casanova, A.~Romero, P.~Li{\`{o}}, and Y.~Bengio, ``Graph attention networks,'' in \emph{6th International Conference on Learning Representations, {ICLR} 2018, Vancouver, BC, Canada, April 30 - May 3, 2018, Conference Track Proceedings}.\hskip 1em plus 0.5em minus 0.4em\relax OpenReview.net, 2018.

\bibitem{tak2021end_gra}
H.~Tak, J.-W. Jung, J.~Patino, M.~Kamble, M.~Todisco, and N.~Evans, ``End-to-end spectro-temporal graph attention networks for speaker verification anti-spoofing and speech deepfake detection,'' in \emph{ASVSPOOF 2021, Automatic Speaker Verification and Spoofing Countermeasures Challenge}.\hskip 1em plus 0.5em minus 0.4em\relax ISCA, 2021, pp. 1--8.

\bibitem{vaswani2017attention}
A.~Vaswani, N.~Shazeer, N.~Parmar, J.~Uszkoreit, L.~Jones, A.~N. Gomez, {\L}.~Kaiser, and I.~Polosukhin, ``Attention is all you need,'' \emph{Advances in neural information processing systems}, vol.~30, 2017.

\bibitem{DBLP:conf/dada/ZhangYTWXF23}
X.~Zhang, J.~Yi, J.~Tao, C.~Wang, L.~Xu, and R.~Fu, ``Adaptive fake audio detection with low-rank model squeezing,'' in \emph{Proceedings of the Workshop on Deepfake Audio Detection and Analysis co-located with 32th International Joint Conference on Artificial Intelligence {(IJCAI} 2023), 2023}, ser. {CEUR} Workshop Proceedings, vol. 3597, 2023, pp. 95--100.

\bibitem{gu2023mamba}
A.~Gu and T.~Dao, ``Mamba: Linear-time sequence modeling with selective state spaces,'' \emph{arXiv preprint arXiv:2312.00752}, 2023.

\bibitem{gu2021combining}
A.~Gu, I.~Johnson, K.~Goel, K.~Saab, T.~Dao, A.~Rudra, and C.~R{\'e}, ``Combining recurrent, convolutional, and continuous-time models with linear state space layers,'' \emph{Advances in neural information processing systems}, vol.~34, pp. 572--585, 2021.

\bibitem{gu2022parameterization}
A.~Gu, K.~Goel, A.~Gupta, and C.~R{\'e}, ``On the parameterization and initialization of diagonal state space models,'' \emph{Advances in Neural Information Processing Systems}, vol.~35, pp. 35\,971--35\,983, 2022.

\bibitem{gupta2022diagonal}
A.~Gupta, A.~Gu, and J.~Berant, ``Diagonal state spaces are as effective as structured state spaces,'' \emph{Advances in Neural Information Processing Systems}, vol.~35, pp. 22\,982--22\,994, 2022.

\bibitem{DBLP:conf/iclr/GuGR22}
\BIBentryALTinterwordspacing
A.~Gu, K.~Goel, and C.~R{\'{e}}, ``Efficiently modeling long sequences with structured state spaces,'' in \emph{The Tenth International Conference on Learning Representations, {ICLR} 2022, Virtual Event, April 25-29, 2022}.\hskip 1em plus 0.5em minus 0.4em\relax OpenReview.net, 2022. [Online]. Available: \url{https://openreview.net/forum?id=uYLFoz1vlAC}
\BIBentrySTDinterwordspacing

\bibitem{ravanelli2018speaker}
M.~Ravanelli and Y.~Bengio, ``Speaker recognition from raw waveform with sincnet,'' in \emph{2018 IEEE spoken language technology workshop (SLT)}.\hskip 1em plus 0.5em minus 0.4em\relax IEEE, 2018, pp. 1021--1028.

\bibitem{liu2017sphereface}
W.~Liu, Y.~Wen, Z.~Yu, M.~Li, B.~Raj, and L.~Song, ``Sphereface: Deep hypersphere embedding for face recognition,'' in \emph{Proceedings of the IEEE conference on computer vision and pattern recognition}, 2017, pp. 212--220.

\bibitem{wang23w_interspeech}
C.~Wang, J.~Yi, J.~Tao, C.~Y. Zhang, S.~Zhang, R.~Fu, and X.~Chen, ``{TO-Rawnet: Improving RawNet with TCN and Orthogonal Regularization for Fake Audio Detection},'' in \emph{Proc. INTERSPEECH 2023}, 2023, pp. 3137--3141.

\bibitem{li23h_interspeech}
J.~Li, Y.~Long, Y.~Li, and D.~Xu, ``Advanced rawnet2 with attention-based channel masking for synthetic speech detection,'' in \emph{Proc. INTERSPEECH 2023}, 2023, pp. 2788--2792.

\bibitem{van2008visualizing}
L.~Van~der Maaten and G.~Hinton, ``Visualizing data using t-sne.'' \emph{Journal of machine learning research}, vol.~9, no.~11, 2008.

\end{thebibliography}

\end{document}